\documentclass[3p,times,twocolumn]{elsarticle}

\usepackage{ecrc}

\volume{00}

\firstpage{1}

\journalname{Nuclear Physics B Proceedings Supplement}

\runauth{}

\jid{nuphbp}

\jnltitlelogo{Nuclear Physics B Proceedings Supplement}

\usepackage{amssymb}

\usepackage[figuresright]{rotating}

\usepackage{graphicx}
\usepackage{subfigure}

\begin{document}

\begin{frontmatter}



\dochead{}

\title{Gamma rays from Galactic Pulsars}


\author[label1]{F. Calore\corref{nota}}
\ead{f.calore$@$uva.nl}

\author[label2,label3]{M. Di Mauro}
\author[label2]{F. Donato}

\address[label1]{GRAPPA Institute, University of Amsterdam, Science Park 904, 1090 GL Amsterdam, The Netherlands}
\address[label2]{Dipartimento di Fisica, Torino University and INFN, Sezione di Torino, Via P. Giuria 1, 10125 Torino, Italy}
\address[label3]{Laboratoire d'Annecy-le-Vieux de Physique Th\'eorique (LAPTh), Univ. de Savoie, CNRS, B.P.110, Annecy-le-Vieux F-74941, France}
\cortext[nota]{Speaker}

\begin{abstract}
Gamma rays from young pulsars and milli-second pulsars are expected to contribute to the diffuse gamma-ray emission measured by the {\it Fermi} Large Area Telescope (LAT) at high latitudes.
We derive the contribution of the pulsars undetected counterpart  by using information from radio to gamma rays and we show that they explain only a  small fraction of the isotropic diffuse gamma-ray background.
\end{abstract}


\end{frontmatter}


\section{The isotropic diffuse gamma-ray background}
Since 5 years, the {\it Fermi} Large Area Telescope (LAT) surveys the gamma-ray 
sky in the GeV energy range. Besides discovering the emission of individual
gamma-ray point-like sources such as blazars, star-forming galaxies and pulsars,
the {\it Fermi}-LAT has confirmed and measured a faint and almost isotropic emission at high latitudes: 
the \emph{isotropic diffuse gamma-ray background} (IGRB) \cite{IDGRB}. 
The relevance of the IGRB is twofold: on the one hand, it is believed to be the superposition of 
several astrophysical contributions, both unresolved sources and diffuse processes; on the other hand,
it can be partly due to the gamma-ray emission from dark matter annihilation in the Galaxy 
and in external galaxies. For this reasons, studying the IGRB is of utmost importance to shed light onto 
the nature of extragalactic and Galactic astrophysical sources as well as to reduce the uncertainty for dark matter
searches by means of this target \cite{Calore:2013yia}. 


In the following, we will focus on the population of pulsars, both young and millisecond (MSP), and on their 
contribution to the IGRB.
Pulsars are indeed the second largest population detected by the LAT. Moreover, MSPs are expected
to reach high latitudes during their evolution and thus they have been proposed to be able to explain 
a significant fraction of the IGRB \cite{2010JCAP...01..005F}.

In Sec.~\ref{sec:pulsars} we summarize the properties of
pulsars and MSP population, their spatial distribution in the Galaxy and their gamma-ray spectral properties
as derived from the Second {\it Fermi}-LAT Catalog of gamma-ray pulsars (2FPC) \cite{2013ApJS..208...17A}.
In Sec.~\ref{sec:simulation} we describe the procedure we follow to generate a mock pulsar and MSP
 population and its gamma-ray emission. We then present our results about the
emission from unresolved young pulsars and MSPs at high- and low-latitudes.
Finally, we draw our conclusions in Sec.~\ref{sec:conclusion}

\section{The pulsar population}
\label{sec:pulsars}
\begin{figure}[t!]
     \begin{centering}
        \subfigure[]{%
            \label{fig:map_res}
            \includegraphics[width=0.45\textwidth]{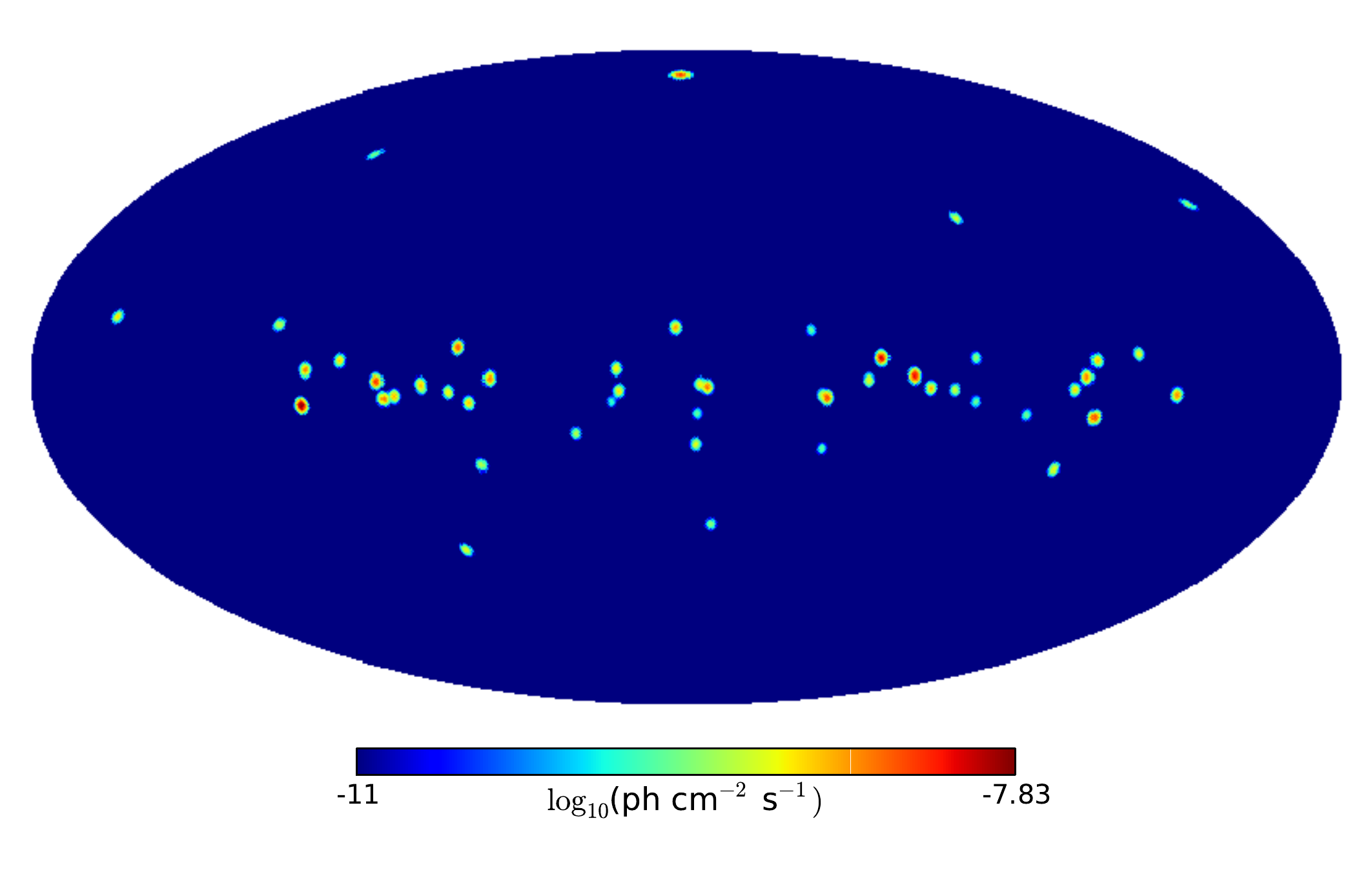}
        } 
        \subfigure[]{%
           \label{fig:map_unres}
           \includegraphics[width=0.45\textwidth]{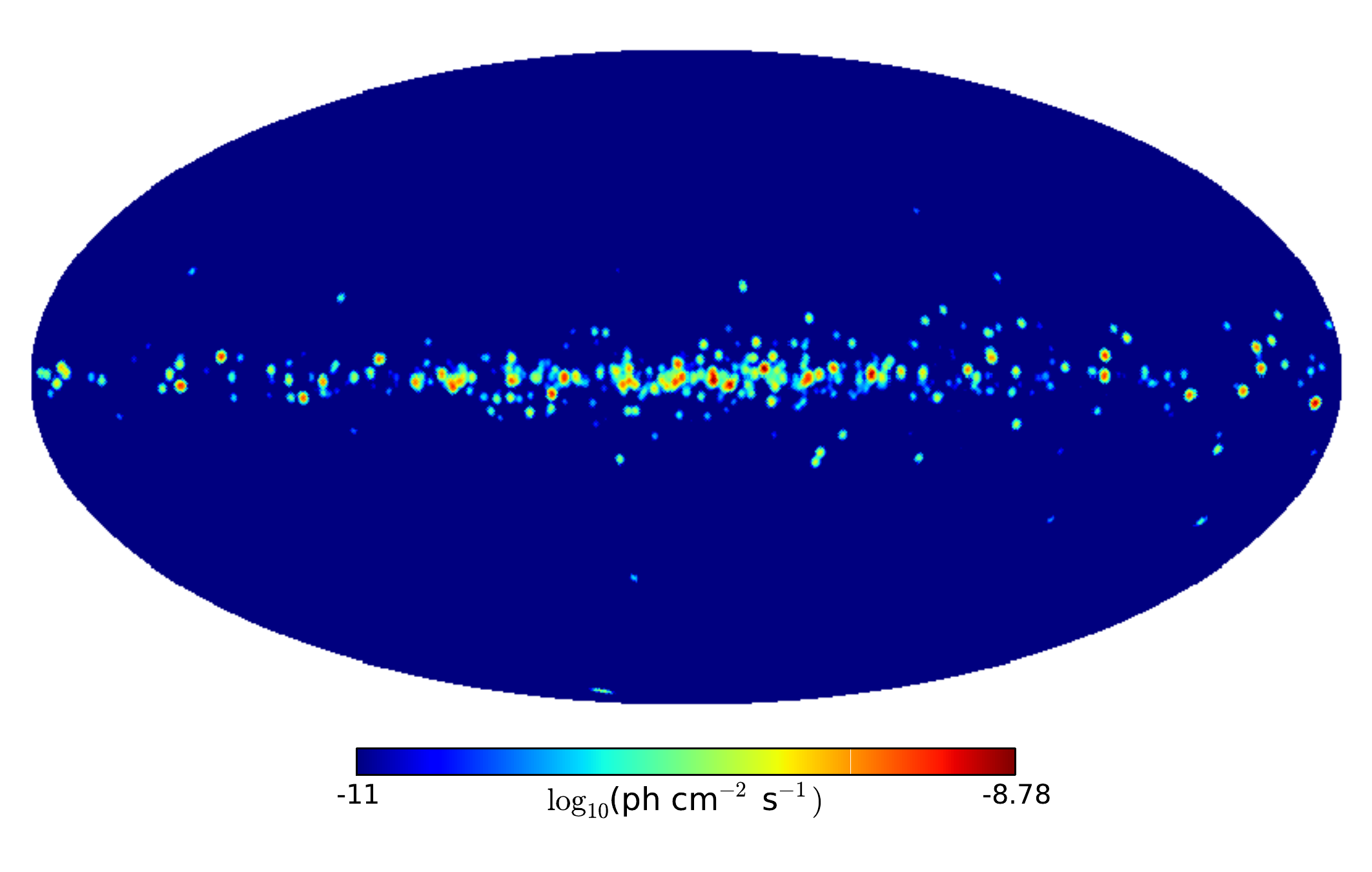}
        }
    \caption{All-sky gamma-ray map of the simulated MSP population. Top: Resolved counterpart. Bottom: Unresolved counterpart.}
   \label{fig:map}
   \end{centering}
\end{figure}

Pulsars are spinning neutron stars with a rotation period of a few milli-seconds up to tens of seconds (separation at P = 15 ms).
The magnetic dipole braking slows down the pulsar rotation period \cite{2000RSPTA.358..831L} and 
the subsequent 
energy loss rate, or spin-down luminosity, is $ \dot{E} = 4 \pi^2 M \dot{P}/P^3$,
where $\dot{P}$ is the period derivative and M is the moment of inertia of the star ($10^{45}$ g cm$^{2}$). The radiation from the star is obtained through the conversion of the spin-down 
luminosity with a given efficiency.

\medskip

In \cite{Calore:2014oga}, we studied the spatial distribution and gamma-ray emission properties of the pulsar and MSP population, using radio and gamma-ray catalogs.
We summarize in the following the main population properties (see~\cite{Calore:2014oga} for a thoroughly discussion and derivation).

The MSP population distribution in the Galaxy was modeled by relying on the Australia National Facility (ATNF)
Pulsar catalog \cite{2005AJ....129.1993M}. We selected 132 MSPs and we determined the distribution of the rotation period P, the surface magnetic field B,
the distance from the Galactic plane z, and the distance from the Galactic center projected on the Galactic plane r. The best-fit distribution parameters are 
summarized in Table 1 of~\cite{Calore:2014oga}.

MSPs are well-known gamma-ray emitters, most likely through curvature radiation. The 2FPC contains 117 gamma-ray pulsars; 40 of them are MSPs. The spectral energy distribution of those sources is parametrized by a power-law with exponential cutoff. In \cite{Calore:2014oga}, we showed that the distribution of the spectral index $\Gamma$ and cutoff energy $E_{\rm cut}$ are compatible with a gaussian function with best-fit parameters: $[\Gamma,\sigma_{\Gamma}]=[1.29, 0.37]$ and $[ \tilde{E}_{\rm{cut}}, \sigma_{\tilde{E}_{\rm{cut}}} ] = [3.38, 0.18]$, where  $\tilde{E}_{\rm{cut}} \equiv \log_{10}(E_{\rm{cut}}/{\rm MeV})$.
The radial and vertical height distribution are consistent to what measured in radio, once the sensitivity of the telescope is properly taken into account.
We adopted a semi-empirical modeling of the gamma-ray emission from MSPs: we assumed that the spin-down luminosity is converted in gamma rays with an efficiency $\eta$, $L_{\gamma} = \eta \dot{E}$. The average value and uncertainty on the parameter $\eta$ were derived by looking at the scatter of data in the $L_{\gamma} - \dot{E}$ plane. To get additional information from gamma-ray data, we computed 95\% C.L.~upper limits on the gamma-ray flux for a sample of 19 sources below the {\it Fermi}-LAT detection threshold. The benchmark value for $\eta$ has been found to be 1 with an uncertainty in the range $\eta = \{0.015, 0.65\}$.

\section{The pulsar gamma-ray sky}
\label{sec:simulation}

We simulated the all-sky gamma-ray emission due to an MSP population modeled as described above.
Each simulated object was identified by its period, magnetic field and position in the Galaxy, as randomly extracted from the corresponding distributions.
 Period and magnetic field determine the spin-down luminosity of the simulated source. By extracting $\eta$ from a uniform distribution in the allowed band, the gamma-ray luminosity can be derived as well as the gamma-ray energetic flux $S_\gamma = L_\gamma / (4 \pi d^2)$.
The energetic flux of each source is used to determine if such an object would be detected or not by the {\it Fermi}-LAT, by comparing the expected $S_\gamma$ from that source with the detection sensitivity curve from the 2FPC (Fig.~17 of \cite{2013ApJS..208...17A}).
In this way, we got a collection of non-detected sources that contribute to the IGRB as diffuse emission, while the number of 
detected objects cannot be larger than the truly observed one.
We assigned to each simulated source a value of $\Gamma$ and $E_{\rm cut}$ extracted from the corresponding distributions. 
The spectral parameters allow to compute the gamma-ray flux that is, in turn, needed to find the spectrum of the source $dN/dE$.
The contribution to the IGRB from the \emph{unresolved} source population, above a given latitude, is given by the sum of the $dN/dE$ of all sources below
the detection threshold.
Figure~\ref{fig:map} shows the resolved and unresolved counterpart of one Monte Carlo realisation.

\label{sec:results}
\begin{figure}[t!]
\begin{centering}
\includegraphics[width=0.45\textwidth]{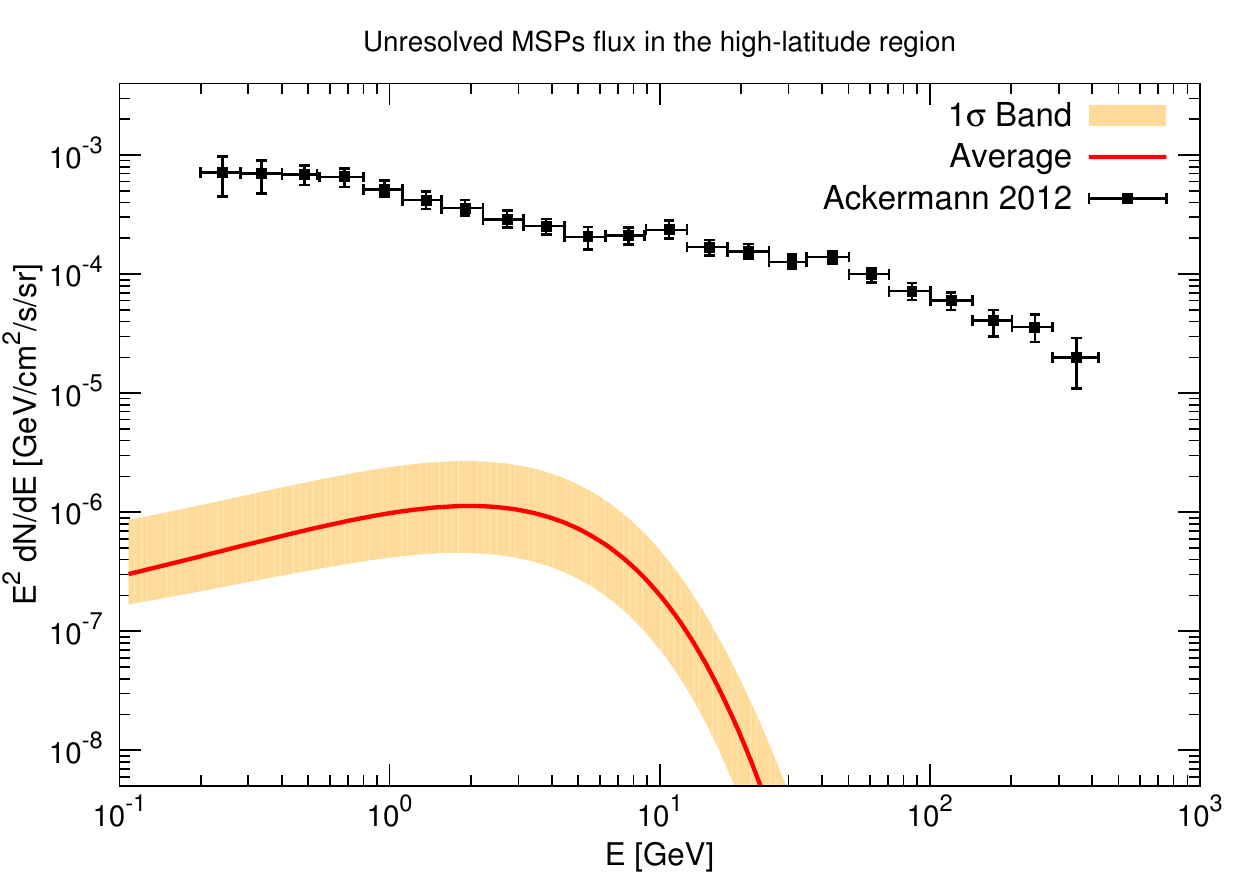}
\caption{Diffuse gamma-ray flux from the MSP unresolved population. The red solid line represents the average over 1000 Monte Carlo realizations, while the light orange band is the 1$\sigma$ uncertainty band.}
\label{fig:BMflux}
\end{centering} 
\end{figure}

The emission from unresolved MSPs at latitudes above 10$^{\circ}$ is shown if Figure~\ref{fig:BMflux}. 
The contribution to the IGRB from unresolved MSPs turns out to be about 0.1\%--0.9\% at the peak (2 GeV) and
about 0.02\%--0.13\% of the integrated IGRB intensity. The uncertainty band is of $\mathcal{O}$(10) at all energies.

We also computed the emission in the inner part of the Galaxy where, recently, several and independent analyses have found an excess above the
standard astrophysical background, the so-called ``{\it Fermi} GeV excess" (as for example~\cite{Macias:2013vya,
Abazajian:2014fta, Daylan:2014rsa, Calore:2014xka}.
We considered two regions: $10^{\circ} \leq |b| \leq 20^{\circ}$ and $l \in [-180^{\circ},180^{\circ}]$, and $|b| \leq 3.5^{\circ}$ and $|l| \leq 3.5^{\circ}$ (the Galactic center region). Beside the MSP contribution, we include the emission of young pulsars (by applying to this population the same methodology outlined above), since pulsars are known to be concentrated more along the disk and might produce a significant emission. In both regions, the contribution from unresolved young pulsars and MSPs might explain only up to 5\%--10\% of the GeV excess. While the spectral properties of young pulsars and MSPs are compatible with the excess, the flux they can produce is not enough to fully explain the signal (see also~\cite{Hooper:2013nhl, Cholis:2014lta}).
Nevertheless, it might possible that a population component associate with the Galactic bulge could explain the intensity and morphology
of the excess \cite{Petrovic:2014xra}.

\section{Conclusion}
\label{sec:conclusion}
We performed a systematic analysis of pulsar and MSP population
properties from radio (ATNF catalog) to gamma rays ({\it Fermi}-LAT).
We demonstrated that MSPs are a marginal component of the IGRB with an uncertainty of 
$\mathcal{O}$(10) at all energies.
MSPs are also a negligible contributor to the gamma-ray
anisotropy signal measured by the {\it Fermi}-LAT, thus indicating that
this should be dominated by other sources.
At low latitudes, the contribution from both young pulsars and
MSPs can explain up to about 10\% of the excess emission
measured in the inner part of the Galaxy.
The MSPs interpretation of the {\it Fermi}-LAT GeV excess is thus 
in tension with spectral and morphological properties of the MSP
disk-like population as we model it from radio and gamma-ray observations.




\bibliographystyle{elsarticle-num}
\bibliography{paper_tmp}







\end{document}